\documentclass[aps,prl,twocolumn,superscriptaddress,showpacs,amssymb,amsmath]{revtex4}
\usepackage{graphicx,color}
\usepackage{color,verbatim}

\newcommand{\bl}{\ensuremath{\lambda}}
\newcommand{\e}{\ensuremath{\varepsilon}}
\newcommand{\g}{\ensuremath{\mathbf{g}}}
\newcommand{\G}{\ensuremath{\mathbf{G}}}
\newcommand{\Ga}{\ensuremath{\mathbf{\Gamma}}}
\newcommand{\Ha}{\ensuremath{\hat{H}}}
\newcommand{\M}{\ensuremath{\mathbf{M}}}
\newcommand{\np}{\ensuremath{\mathcal{N}_0}}
\newcommand{\om}{\ensuremath{\omega}}
\newcommand{\Si}{\ensuremath{\mathbf{\Sigma}}}
\newcommand{\T}{\ensuremath{\mathcal{T}}}
\newcommand{\bV}{\ensuremath{\bar{V}}}

\begin{document}

\title{Phonon-assisted current noise in molecular junctions}

\author{Federica Haupt}
\affiliation{Fachbereich Physik, Universit\"at Konstanz, D- 78457
Konstanz, Germany}

\author{Tom\'{a}\v{s} Novotn\'y} 
\affiliation{Department of Condensed Matter Physics, Faculty of
Mathematics and Physics, Charles University in Prague, Ke Karlovu 5,
CZ-121 16 Praha 2, Czech Republic}

\author{Wolfgang Belzig}
\affiliation{Fachbereich Physik, Universit\"at Konstanz, D- 78457
Konstanz, Germany}

\begin{abstract}
We investigate the effects of phonon scattering on the electronic
current noise through nanojunctions using the non-equilibrium
Green's functions formalism extended to include the counting field.
In the case of weak electron-phonon coupling and a single broad
electronic level we derive an analytic expression for the current
noise at arbitrary temperature and identify physically distinct
contributions based on their voltage dependence. We apply our theory
to the experimentally relevant case of a D$_2$ molecule placed in a
break-junction and predict a significant inelastic contribution to
the current noise.
\end{abstract}

\date{\today}
 \maketitle

\noindent
\emph{Introduction.}--- The fabrication of atomically sharp contacts
has opened up the possibility of creating junctions formed by a
single molecule bridging metal electrodes~\cite{Cuniberti}. However,
the inherent complexity of this emerging field poses fundamental
challenges and makes junctions formed by very simple molecules (e.g.
hydrogen~\cite{Smit,Djukic_PRB,Djukic} or water~\cite{Tal}) an
invaluable testbed from both the experimental and the theoretical
point of view. In these systems, inelastic effects due to the
interaction between transport electrons and molecular vibrational
modes (phonons) result in an abrupt change of the differential
conductance at the onset of the phonon emission. These features have
been exploited to establish unambiguously the presence of the
molecule in the contact~\cite{Smit} and, when combined with
shot-noise measurements~\cite{Djukic,Kiguchi}, they allow for a
detailed characterization of the junction.

Theoretical descriptions of inelastic transport through
nanojunctions have so far focused mainly on the current-voltage
characteristics, while less attention has been paid to the study of
noise. Phonon-scattering effects on the differential conductance
have been addressed both with ab-initio
methods~\cite{Frederiksen:PRL04,Paulsson:RapCom,Viljas,delaVega,Frederiksen:PRB07,Asai}
and with simplified (one-level) models~\cite{Mitra, Koch, Ryndyk,
Egger}. Noise calculations based on one-level models have also been
put forward within the rate equation approach~\cite{Mitra, Koch} or
within the non-equilibrium Green's functions (NGF) formalism
~\cite{Galperin:PRB06} with a mean-field-like approximation for the
noise. 
In this work we study inelastic effects on the
current noise with the NGF approach taking consistently into account
all the correlations due to phonon-assisted scattering up to a given
order in the $e$-ph interaction. We apply our theory to the case of
a hydrogen-bridge junction and predict a significant inelastic
contribution to the current noise.


\emph{Model \& methods.}--- The system we consider can be
 represented as a central device region which is
tunnel-coupled to non-interacting metallic leads
$\Ha=\Ha_C+\Ha_{L,R}+\Ha_T$. Neglecting the spin
degree of freedom, the central region can be described by the
Hamiltonian $\Ha_C=\Ha_0+\sum_{\ell}\hbar
\om_{\ell}\hat{b}_{\ell}^{\dag}\hat{b}_{\ell}+\sum_{\ell}\sum_{i,j}M_{\ell}^{ij}\hat{d}^{\dag}_i\hat{d}_j(\hat{b}_{\ell}^{\dag}+\hat{b}_{\ell})$,
where $\hat{d}^{\dag}_i$ and $\hat{b}^{\dag}_{\ell}$ are the electron and phonon creation operators,
$\Ha_{0}=\sum_{i,j}H_0^{ij}\hat{d}^{\dag}_{i}\hat{d}_j$ is the
single-particle effective Hamiltonian of the electrons moving in a
static arrangement of atomic nuclei, and $\M_{\ell}$ is the $e$-ph
coupling matrix for the ${\ell}$-th phonon mode. Here, boldface
notation stands for matrix over electronic space.  The leads and
tunneling Hamiltonians are given by
$\Ha_{L,R}=\sum_{k,\alpha=L,R}\varepsilon_{\alpha,k}\hat{c}^{\dag}_{\alpha,k}\hat{c}_{\alpha,k}$
and
$\Ha_T=\sum_{k,\alpha=L,R}(V_{\alpha,k}^i\hat{c}^{\dag}_{\alpha,k}\hat{d}_i^{\phantom{\dagger}}+
h.c.)$. The states in the leads are occupied according to the Fermi
distribution $f_{\alpha}(\e)=[1+e^{\beta (\e-\mu_{\alpha})}]^{-1}$,
where $\beta=1/k_B T$ is the inverse temperature and $\mu_ {\alpha}$
is lead-$\alpha$ chemical potential, with $\mu_L-\mu_R=eV$ fixed by
the applied bias voltage.

To calculate the current and the noise we employ the extended NGF
technique~\cite{Nazarov} to find
the cumulant generating function
$\mathcal{S}(\lambda)=I/(ie)\lambda+S/(2e^{2})\lambda^{2}+\cdots$,
from which the current $I$ and the noise $S$ can be calculated
straightforwardly. The key idea~\cite{Levitov} is to modify the
Hamiltonian by adding a time-dependent phase $\lambda(t)/2$ to the
tunneling matrix elements $V^{i}_{L,k}$, with $\bl(t)=\pm\lambda$ for
$t$ on the upper and the lower Keldysh contour, respectively.  It
has been shown in Ref.~\cite{GogolinKomnik} for the Anderson
model, that $\partial_{\lambda}\mathcal{S}(\lambda)$ is related to the
Keldysh Green's function $G_{\bl}(t,t^{\prime})=-i\langle {\cal
  T_{C}} \hat{d}(t) \hat{d}^{\dagger}(t^{\prime})\rangle_{\lambda}$, where the
expectation value is now evaluated in the presence of $\bl(t)$.
Generalizing that result to the multilevel case, we write
\begin{equation}
\frac{\partial{\mathcal{S}}( \bl)}{\partial{\lambda}}\!= \!\int\!
\frac{d \e}{2 \pi \hbar} {\rm
Tr}\{{\Ga_L}[e^{-i\bl}(1-f_L)\G_{\bl}^{-+}+e^{i\bl}f_L
\G_{\bl}^{+-}]\}.
\end{equation}
Here $\Gamma_{\alpha}^{ij}(\e)=2 \pi \sum_k V^i_{\alpha,k}
{V^{{j}^*}_{\alpha,k}}\delta(\e-\e_{k,\alpha})$ is the level
broadening due to the coupling to lead-$\alpha$ and
$\G_{\bl}^{-+}$, $\G_{\bl}^{+-}$  are appropriate components of the
solution of the non-equilibrium Dyson equation
$\check{\G}_{\lambda}={\check{\g}}_{\lambda}+{\check{\g}}_{\lambda}\check{\Si}_{\lambda}{\check{\G}}_{\lambda}
$. The check sign indicates matrices in the Keldysh space and
the superscripts $-/+$ correspond to the forward/backward branch of
the Keldysh-contour. The matrix $\check{\g}_{\bl}$ is Green's
function of the system in the presence of the leads and of the
counting field but without the {\em e}-ph interaction. Its inverse
is given by
\begin{widetext}
\begin{equation}
  \check{\g}_{\bl}^{-1}(\e)=\left( \begin{array}{cc} \e\,\mathbf{1}-\mathbf{H}_0-i\sum_{\alpha}\Ga_{\alpha}[f_{\alpha}(\e)-1/2]&
                                    i\Ga_Le^{i\bl}f_L(\e)+i \Ga_R f_R(\e)\\
                                    -i \Ga_Le^{-i\bl}[1-f_L(\e)]-i\Ga_R[1-f_R(\e)]&
                                    -\e\,\mathbf{1} +\mathbf{H}_0-i\sum_{\alpha}\Ga_{\alpha}[f_{\alpha}(\e)-1/2] \end{array} \right).
\end{equation}
\end{widetext}
As $\check{\g}_{\bl}$ already includes the coupling to the
leads, $\check{\Si}_{\lambda}$ is  the self-energy solely due to the
{\em e}-ph coupling. Being interested in the weak coupling limit, we
expand the Dyson equation to the lowest (second) order  in the {\em
e}-ph coupling
$\check{\G}_{\lambda}\approx{\check{\g}}_{\lambda}+{\check{\g}}_{\lambda}\check{\Si}_{\lambda}^{(2)}{\check{\g}}_{\lambda}$,
where ${{\Si}_{\lambda}^{(2)}}$ is given by the Fock diagram
$(\eta,\bar{\eta}=\pm)$
 \begin{equation}\label{eq:Sigma2}
{{\Si}_{\lambda}^{(2)}}^{\eta\bar{\eta}}(\e)=i\sum_{\ell}
\int\frac{d\e'}{2\pi}\,d^{\eta\bar{\eta}}_{\ell}(\e-\e')\M_{\ell} \,
\g^{\eta\bar{\eta}}_{\bl}(\e')\M_{\ell}.
\end{equation}
The Hartree term has been  neglected since it cannot
contribute by any truly dynamical features in which we are primarily
interested 
Above, $d_{\ell
}^{\eta\bar{\eta}}(\e)$ stand for {\em free} phonon Green's
functions for the $\ell$-th phonon mode $ d_{\ell
}^{\pm\pm}(\e)=\sum_{s=\pm}\big[-i\pi (2\mathcal{N}_{\ell
}+1)\delta(\e+s\hbar\om_{\ell })\pm
\mathcal{P}\frac{s}{\e+s\hbar\om_{\ell }} \big]$ and $d_{\ell
}^{\mp,\pm}(\e)=-2\pi i [(\mathcal{N}_{\ell }+1)\delta(\e \pm
\hbar\om_{\ell })+\mathcal{N}_{\ell }\delta(\e \mp \hbar \om_{\ell
})]$, with $\mathcal{N}_{\ell }$ the (generally non-equilibrium)
occupation of mode $\ell$.

Truncating the Dyson equation to the second order in ${\mathbf
M}_{\ell}$ directly yields the  expressions for $\check{\G}_{\bl=0}$
and  $\partial_{\bl}\check{\G}_{\bl}\big|_{\bl=0}$, which are the
ingredients to evaluate the current and the noise.
Integration over energy can be performed analytically assuming
the electronic structure to be slowly changing over few multiples of
a typical phonon energy around the Fermi level $E_F$ and
approximating $\Ga_{\alpha}(\e)\approx\Ga_{\alpha}(E_F)$ and
$\g^{r}_{\bl=0}(\e)\approx\g^{r}_{\bl=0}(E_F)$~\cite{Paulsson:RapCom,Frederiksen:PRB07,
Viljas,delaVega}. 

Physically important non-equilibrium phonon heating
effects \cite{MadsPreprint} can be qualitatively taken into account
by  a rate equation for the average phonon occupation number
$\mathcal{N}_{\ell}$~\cite{Frederiksen:PRL04}, which can be viewed
as a kinetic-equation-like approximation to the full NGF studies
\cite{Ryndyk,Asai}. In the broad-level approximation introduced
above, this leads to a bias-dependent occupation number
$\mathcal{N}_{\ell}(V)=n_{B}(\om_{\ell})+\alpha_d n_{\ell}(V)$,
where $n_B(\om)=1/(e^{\beta \hbar \om}-1)$ is the Bose distribution,
$\alpha_d$ is a parameter which depends on strength of the external
phonon damping, and $n_{\ell}(V)$ takes into account the power
dissipated by the transport electrons into the phonon mode~\cite{Paulsson:RapCom}. 
In the following we will focus on the two opposite regimes of (i)
thermally equilibrated phonons ($\alpha_d=0$), and (ii) non-equilibrated 
phonons ($\alpha_d=1$). In this case  
it is $n_{\ell}(V)\approx (|eV|/\hbar \omega_{\ell}-1)\theta(|eV|-\hbar \omega_{\ell})/4$ 
for $k_B T \ll \hbar \omega_{\ell} $~\cite{Frederiksen:PRB07}.


\emph{Analysis.}--- For sake of clarity, we focus here
only on the case of a single electronic level $\e_0$ with symmetric
coupling to both leads $\Gamma_L=\Gamma_R=\Gamma$ and coupled to a
single phonon mode with frequency $\om_0$, occupation $\np$ and
coupling constant $M$. Already such a simple model reveals many
essential features of inelastic transport through nano-junctions
such as the phonon-induced step behavior of the differential
conductance~\cite{Tal, Egger}. Before focusing on the inelastic
corrections to the noise, we shortly reexamine the results for the
current. We find that the
current through the device is given by $I=I_{\rm el}+I_{\rm inel}$,
with $I_{\rm el}=(e^2/h) \times \T V$ and
\begin{equation}\label{eq:I_inel}
\begin{split}
I_{\rm inel}&=\frac{e\, \gamma_{e\rm ph} \om_0}{ 2 \pi}\Big[(1-2\T)
\frac{W(\bV-1)-W(\bV+1)}{2}\\
&+(2 \np+1)(3-4\T)\bV \Big],
\end{split}
\end{equation}
with the reduced voltage $\bV=eV/\hbar\om_0$, the dimensionless
$e$-ph coupling $\gamma_{e\rm ph}={M^2\T^2}/{\Gamma^2}$, and $W(x)=x
\coth(\beta \hbar \om_0 \,x/2)$. Here $\T$ is the elastic
transmission coefficient $\T=|G^{r}|^2\Gamma^2=\Gamma^2/(\Delta^2
+\Gamma^2)$ and $\Delta=(E_F-\e_0)$ gives the position of the single
level with respect to the Fermi energy. The inelastic current
results then from the sum of two contributions with distinct
behavior with respect to the bias voltage: while the first term of
Eq. \eqref{eq:I_inel}, which is responsible for the step features in
the non-linear conductance, saturates to constant values for
$|\bV|>1$, the second one grows linearly with $V$ for
$\np=n_B(\om_0)$ and quadratically in the case of non-equilibrated
phonons. This second term has a clear physical interpretation in
terms of electrons experiencing the coupling to the phonon
as a stochastic quasi-static shift in the energy of the
level. This in turn affects the transmission coefficient, which
becomes dependent on the displacement of the oscillator
$\T(Q)$. Averaging over $Q$ and retaining only terms to the second
order in $M$ one obtains
\begin{equation}\label{eq:T(Q)}
\langle \T(Q) \rangle\!=\!\left\langle \!\frac{\Gamma^2}{(\Delta
-MQ)^2+\Gamma^2}\!\right\rangle \!\approx \! \T +\gamma_{e\rm
ph}(3-4\T)\langle Q^2\rangle,
\end{equation}
where $\langle \cdot \rangle$ indicates the average over a
stationary distribution of the oscillator, so that $\langle
Q\rangle=0$ and $\langle Q^2\rangle=(2 \np +1)$. We can therefore
interpret the last term of Eq.~\eqref{eq:I_inel} in terms of an
elastic-like contribution with averaged transmission over the
fluctuating position of the oscillator
\footnote{Eq.~\eqref{eq:I_inel} contains an additional term
$(e\gamma_{e\rm ph}/h)\times 2(1+2\np)(1-\T)eV$ with respect to
Refs.~\cite{Paulsson:RapCom,Frederiksen:PRB07}. This term gives a
finite inelastic contributions to the current also below the phonon emission threshold.}

\begin{figure}
\begin{center}
{\resizebox{0.8\columnwidth}{!}{\includegraphics{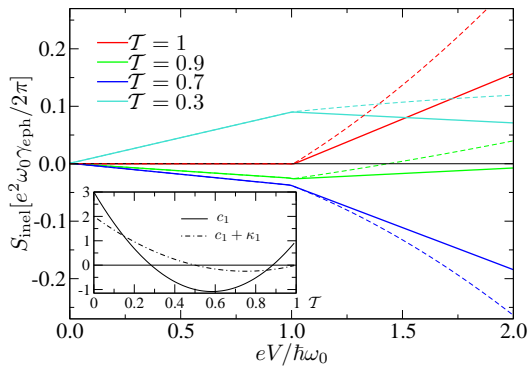}}}
\end{center}
\caption{(Color online) Inelastic noise $S_{\rm inel}$ as a function
of the bias voltage at zero temperature $T=0$ for different values of the transmission coefficient.  Both cases of
equilibrated ($\alpha_d=0$, thick lines) and non-equilibrated
phonons ($\alpha_d=1$, dashed lines) are shown. 
Inset: Plot of $c_1|_{\{T=0, |eV|=\hbar \om_0\}}$ (full line) and
$(c_1+\kappa_1)|_{\{T=0, eV=0\}}$ (dash-dotted line) as a function of
$\T$ ($\np=0$ for these parameter values). }
\label{fig:SinelT0}
\end{figure}

We now turn to the main result of our work which is the
phonon-assisted noise. The current noise through the device is given
by $S=S_{\rm el}+S_{\rm inel}$, with the standard expression for the
elastic noise $S_{\rm el}=(e^2/h) \{\T^2/\beta+\T (1-\T)\hbar\om_0
W(\bV) \} $~\cite{Blanter} and with
\begin{equation} \label{eq:S_inel}
\begin{split}
&S_{\rm inel}=\frac{e^2 \gamma_{e\rm ph}\om_0}{ 2\pi}\biggl\{\Big[\frac{c_0}{\beta \hbar \om_0}+c_1\,W(\bV)\Big]  +\kappa_0 \\
& - \kappa_1 \Big[\sum_{s=\pm 1}\!\frac{W(\bV+s)}{2} - W(\bV)\Big]
-\!\frac{\kappa_2}{\beta \hbar\om_0}\!\sum_{s=\pm 1}\!s W'(\bV+s) \biggr\}
\end{split}
\end{equation}
being the correction due to inelastic scattering. Here $W'(x)=dW/d
x$ and the coefficients read $c_0=4(2\np+1)\T(5-6\T)$,
$c_1=(2\np+1)(12\T^2-14\T+3)$, $\kappa_0=4\T (1-\T) (1+2 \np)\{[1+2
n_B(\om_0)]-2/\beta \hbar\om_0\}- (1-2\T)^2$,
$\kappa_1=4\T(1-\T)[1+2 n_B(\om_0)]- (1-2\T)^2(1+2\np)$, and,
finally, $\kappa_2=2\T(1-2\T)$. In the zero bias limit $S_{\rm
inel}$ satisfies the fluctuation-dissipation theorem, i.e.  $S_{\rm
inel}|_{V=0}=2\mathcal{G}_{\rm inel}/\beta$, where
$\mathcal{G}_{\rm inel}=(\partial I_{\rm inel}/\partial V)|_{V=0}$
is the inelastic correction to the linear conductance.

\begin{figure}
\begin{center}
{\resizebox{0.8\columnwidth}{!}{\includegraphics{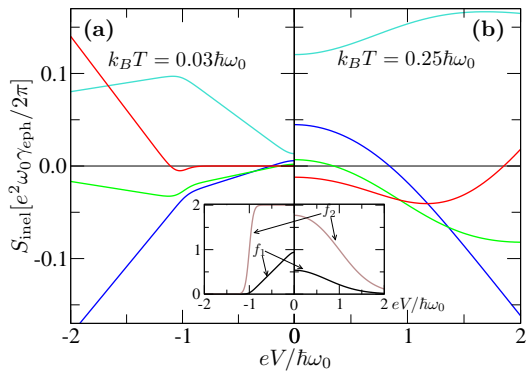}}}
\end{center}
\caption{(Color online) Temperature dependence of inelastic noise.
{\bf (a)}
$S_{\rm inel}$ for the case of equilibrated phonons ($\alpha_d=0$)
at $k_B T=0.03\,\hbar\om_0$ (a typical experimental value
~\cite{Paulsson:RapCom,Djukic_PRB}). Different lines correspond to
various values of the transmission coefficient $\T$ (color code
identical to Fig. \ref{fig:SinelT0}). \textbf{(b)} Same as in (a) but
at higher temperature $k_B T=0.25\,\hbar\om_0$.  Inset:  Plots of $f_1(\bV)=\frac{1}{2}[W(\bV+1)-2
W(\bV)+W(\bV-1)]$ and $f_2(\bV)=W'(\bV+1)-W'(\bV-1)$ at $k_B T=0.03
\,\hbar\om_0$ (left panel) and $k_B T=0.25
\,\hbar\om_0$ (right panel). These functions characterize the dynamic contributions to
$S_{\rm inel}$ in Eq.~\eqref{eq:S_inel}.} \label{fig:SinelT}
\end{figure}

Similarly as for the current, we interpret the first term of
Eq.~\eqref{eq:S_inel} as a quasi-static correction to the elastic
noise due to averaged transmission over the displacement of the
oscillator. In fact, along the same line which lead to
Eq.~\eqref{eq:T(Q)}, it is easy to see that $\gamma_{e \rm ph}c_0$
and $\gamma_{e \rm ph}c_1$ correspond exactly to the contribution of
order $M^2$ to $\langle T(Q)^2\rangle$ and $\langle
T(Q)(1-T(Q))\rangle$, respectively. The remaining terms are dynamic
contributions which take into account the phonon exchange effects.
Analogously to the current, the quasi-static contribution has a
distinct voltage behavior compared to the dynamic one for large
voltage
, being the only one which does not saturate for $|\bV|> 1$.

In the limit of zero temperature Eq.~\eqref{eq:S_inel} simplifies
noticeably  becoming
\begin{equation}
\begin{split}
S_{\rm inel}|_{T=0}&=\frac{e^2 \gamma_{e\rm ph}\om_0}{2\pi}\big\{\kappa_0|_{T=0}+ c_1|_{T=0}\cdot|\bV|\\
&-\kappa_1|_{T=0}\cdot(1-|\bV|)\theta(1-|\bV|) \big\}.
\end{split}
\end{equation}
The inelastic noise is then a piecewise function characterized by
the  coefficient $c_1|_{T=0}$ for $|\bV|>1$ and by the combination
$(c_1+\kappa_1)|_{T=0}$ for $|\bV|<1$, which takes into account the
competition between quasi-static contribution and the dynamic one.
The signs of these coefficients determine whether $S_{\rm inel}$ is
an increasing or decreasing function of voltage, leading to the rich
behavior presented in Fig.~\ref{fig:SinelT0}. Interestingly,
depending on the value of  $\T$ both $(c_1+\kappa_1)|_{T=0}$ and
$c_1|_{T=0}$ can be negative, resulting in a negative contribution
to the inelastic noise. In other words, for a rather wide range of
transmissions, phonon scattering events lead to suppression of the
current noise through the device. Note that both $c_1$ and
$\kappa_1$ may depend on voltage via $\np$. However, since the
phonon heating becomes effective only above the phonon emission
threshold, $S_{\rm inel}$ is always a linear function for $|\bV|<1$,
while energy accumulation into the phonon mode results in the
quadratic increase of $S_{\rm inel}$ for $|\bV|>1$
in the case of no external damping (non-equilibrated phonons, $\alpha_d=1$).

For finite temperatures qualitatively new features appear (see
Fig.~\ref{fig:SinelT}). For small temperatures the curves get
rounded around $\bar V=1$ and for high transmission even an
additional dip occurs (see Fig.~\ref{fig:SinelT}a). The changes
become more pronounced if the temperature is of the order of the
phonon frequency, when the kinks of $S_{\rm inel}$ are largely
washed out. Nevertheless, it is still possible for a wide set of
parameters to have a negative inelastic correction to the noise and
even a sign change at some finite $|\bar V|$ (see
Fig.~\ref{fig:SinelT}b).

\emph{Results.}--- We apply now our formulas to the case of a single
hydrogen molecule between platinum atomic contacts~\cite{Smit}.
Experimentally, it has been shown that hydrogen can form a stable
bridge between Pt electrodes with conductance close to the quantum
unit~\cite{Smit,Djukic_PRB} carried dominantly by a single, almost
transparent channel~\cite{Djukic}. Such a picture has been confirmed
by first principle calculations showing that a single conductance
channel forms due to strong hybridization between the H$_2$
anti-bonding state and the Pt metal states, while the bonding state
is not involved in the transport~\cite{Thygesen}.

\begin{figure}
\begin{center}
{\resizebox{0.8\columnwidth}{!}{\includegraphics{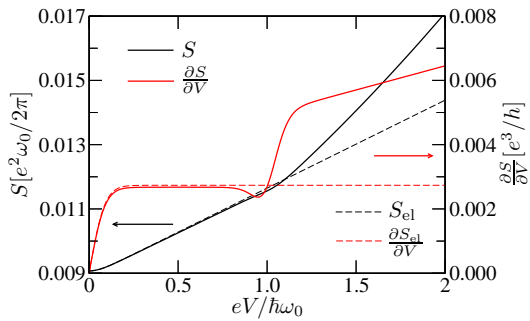}}}
\end{center}
\caption{(Color online) Total current noise $S=S_{\rm el}+S_{\rm
    inel}$ through a D$_2$ molecule and its
  derivative $\partial S/\partial V$  as a function of
  voltage. The elastic contributions $S_{\rm el}$ and $\partial S_{\rm el}/\partial V$ are plotted as dashed lines for comparison.
  Parameter values $k_BT=0.029\,\hbar\omega_0$, $\hbar
  \omega_0=50 {\rm m}eV $, $\tau=0.9825$, $\gamma_{e\rm ph}=0.011$ and
  $\alpha_d=1$ are taken from Ref.~\cite{Paulsson:RapCom} and
  correspond to the conductance measurements in
  Ref.~\cite{Djukic_PRB}. The noise level at $eV=\hbar\omega_{0}=50$
  meV corresponds to experimentally accessible $3\times 10^{-27} {\rm
    A}^2/{\rm Hz}$. Phonon heating ($\alpha_d=1$) responsible for the
  finite slope of $\partial S/\partial V$ for $|eV|>\hbar\om_0$ must
  be included because of the large mass mismatch between the D$_2$
  molecule and Pt atoms.}
\label{fig:noiseD2}
\end{figure}

Fig.~\ref{fig:noiseD2} represents our prediction for the
phonon-assisted noise through a D$_2$ junction, where typical values
for $\T$, $k_B T$ and $\gamma_{e \rm ph}$ have been taken from
Ref.~\cite{Paulsson:RapCom} and correspond to the experimental data
of Ref.~\cite{Djukic_PRB}. One main message of our work is that,
despite of the very weak $e$-ph coupling, inelastic corrections give
a sizable contribution to the total noise through a D$_2$ junction
for $|eV|>\hbar\om_0 $. On the other hand, inelastic corrections are
negligible for $|eV|<\hbar\om_0 $, thus justifying the
interpretation of noise measurement in this regime in
terms of elastic theory (as done in~\cite{Djukic}).

In conclusion, we have presented a perturbative scheme for the
calculation of the inelastic contribution to the current noise in
systems with weak $e$-ph interaction. In the experimentally relevant
case of a single broad level, we have derived an analytic expression
for the inelastic noise at arbitrary temperature and distinguished
terms that correspond to simple renormalization of the transmission
coefficient from those which contain true dynamical effects.
Applying our theory to the case of a D$_2$ junction, we predict a
sizable contribution to the total noise due to inelastic processes.
Our scheme can be straightforwardly extended
beyond the present model to  cases with multiple
electronic levels and phonon modes, asymmetric coupling to leads,
energy-dependent transmission, and/or moderate $e$-ph coupling with
application in current ab-initio methods~\cite{Paulsson:RapCom,Viljas,delaVega,
Frederiksen:PRB07}.
 
We thank M.~Brandbyge, J.~M.~van~Ruitenbeek, and Y.~Utsumi for
useful discussions. We acknowledge the financial support by DFG via
SFB 767 (F.~H. and W.~B.), by the Czech Science Foundation via the
grant 202/07/J051 and the Ministry of Education of the Czech
Republic via the research plan MSM 0021620834 (T.~N.).

{\em Note added:} After submission of the present
manuscript two related works 
were published~\cite{Schmidt}.


\end{document}